\documentclass[10pt]{article}

\usepackage{amsmath}
\usepackage{amsfonts}
\usepackage{amssymb}

\usepackage[dvips,colorlinks=true,urlcolor=blue,linkcolor=blue,citecolor=blue]{hyperref}

\newtheorem{theorem}{Theorem}

\newtheorem{definition}{Definition}

\newenvironment{proof}[1][Proof]
    {\begin{trivlist}\item[\hskip \labelsep \textit{#1.}]}
    {\hfill$\square$ \end{trivlist}}

\newcommand{\rmi}{\ensuremath{\textrm{i}}}
\newcommand{\rmd}{\ensuremath{\textrm{d}}}

\begin{document}

\title{Times
of arrival: Bohm beats Kijowski}
\author{M Ruggenthaler\thanks{Present Address: Institut f\"ur Medizinische Physik, Medizinische
Universit\"at Innsbruck, M\"ullerstr. 44, A-6020 Innsbruck, AUT},
G Gr\"ubl and S Kreidl\\\footnotesize Institut f\"ur Theoretische
Physik, Universit\"at Innsbruck\\[-5pt]\footnotesize Technikerstr. 25,
A-6020 Innsbruck, AUT\\\small E-mail: \footnotesize
\href{mailto:gebhard.gruebl@uibk.ac.at}{Gebhard.Gruebl@uibk.ac.at}
}
\date{}

\maketitle

\begin{abstract} We prove that the Bohmian arrival time of the 1D Schr\"{o}dinger
evolution violates the quadratic form structure on which
Kijowski's axiomatic treatment of arrival times is based. Within
Kijowski's framework, for a free right moving wave packet $\Psi$,
the various notions of arrival time (at a fixed point $x$ on the
real line) all yield the same average arrival time
$\overline{t}_{K_{ij}}\left(  \Psi\right)  $. We derive the
inequality $\overline{t}_{B}\left(  \Psi\right)
\leq\overline{t}_{K_{ij}}\left( \Psi\right)  $ relating the
average Bohmian arrival time to the one of Kijowksi. We prove that
$\overline{t}_{B}\left( \Psi\right)  < \overline{t}_{K_{ij}}\left(
\Psi\right)  $ if and only if $\Psi$ leads to position probability
backflow through $x$.\\\\
PACS numbers: 03.65
\end{abstract}

\section{Introduction}
Let a ready particle detector be exposed to a propagating one
particle wave function. What is the probability distribution of
the time when the detector clicks? Even in the simplest case of a
free one dimensional Schr\"odinger wave function, the proposed
answers to this question for an "intrinsic, free arrival time
distribution" remain controversial. (See e.g. the introduction to
\cite{TOA in QM}.) The problem arises from the fact that quantum
mechanics provides probability distributions only for the outcomes
of measurements performed at a certain time $t$, which has to be
chosen by the observer. And no such choice shows up in the above
situation.

Among the various notions of arrival time, offered by standard
quantum mechanics, the most prominent one arises from the
generalized resolution of the identity associated with the arrival
time operator of Aharonov and Bohm \cite{A-B-TOA}. This operator's
density of arrival times also belongs to a set of arrival time
densities proposed by Kijowski \cite{kijowski} and it is unique
within this set insofar as it minimizes, for every wave function,
the variance of arrival times. Kijowski determined his set from a
list of axiomatic properties that seem plausible within standard
quantum mechanics. A summary of these matters is given by
Egusquiza et al. in section 10 of \cite{TOA in QM}.

Bohmian mechanics extracts a probability space of particle
trajectories from each solution of a configuration space
Schr\"odinger equation. This trajectory space seems a natural
candidate from which to derive the intrinsic arrival time
distribution of an arbitrary wave function. One simply has to
answer the following question for any $t\in\mathbb{R}$: What is
the probability measure of the subset of trajectories which
intersect the detector's volume at any time $s$ prior to time $t$?
Leavens seems to have made the first use of this idea \cite{LEA1}.
For certain 1D scattering wave functions $\psi_{t}$, he derived
the position probability current $J_{x}(\psi_{t})$ at point $x$ to
be identical (up to normalization) with the conditional
probability density for the arrival at point $x$ at time $t$. (The
conditioning is made to the event that an arrival at $x$ occurs at
all.) In the 3D case the Bohmian strategy has been outlined by
Daumer, D\"{u}rr, Goldstein, and Zanghi in their contribution to
\cite{FMV} and in \cite{DDGZ2}. Later on for the general 1D case
Leavens has argued that within Bohmian mechanics $\left\vert
J_{x}(\psi_{t})\right\vert $ (up to normalization) is identical
with the conditional probability density for the arrival at point
$x$ at time $t$ \cite{LEA2}. While Leavens' argument is correct
under certain limited circumstances it is wrong in general. A
cutoff procedure for reentering trajectories is missing from
$\left\vert J_{x}(\psi_{t})\right\vert $ \cite{DDGZ2,gr}. Instead
of this, as has been shown in \cite{kreidl+GG}, a more complicated
expression, involving the current $J_{x}(\psi_{s}) $ at all times
$s$ prior to $t$, yields, within Bohmian mechanics, the
conditional probability density for the arrival at point $x$ at
time $t$.

In the present work we study the question whether the Bohmian
arrival time density, restricted to free 1D positive momentum wave
functions, belongs to the set of arrival time densities introduced
axiomatically by Kijowski. We shall prove that it does not do so,
since already the basic quadratic form structure, which Kijowski
assumes, is violated. No wonder that the expectation values of
arrival times according to Bohm and according to Kijowski in
general differ. We shall show that the Bohmian expectation value
is less or equal to the one according to Kijowski. And it is
exactly for wave functions without position space probability
backflow through the arrival point $x$ that the two expectation
values coincide.

This leads us to the question of measurability of Bohmian arrival
times. As we understand it, the main virtue of Bohmian mechanics
with its introduction of a definite position in configuration
space, is the fact that it provides the mathematical structure to
represent within quantum theory the empirical fact that individual
systems have properties. In this manner Bohmian mechanics gets rid
of the quantum measurement problem. But it does so only if it is
assumed that a system's properties, which may encompass an
observer's perception, are completely determined by its Bohmian
configuration. (Unlike wave functions, Bohmian positions are
definite and unsplit.) Therefore it seems likely that a detection
event happens as soon as a sufficient change in the detector's (or
the observers) Bohmian configuration has taken place. This happens
at about the instant, when the Bohmian position of the detected
particle passes the detector. Why? Because the 'empty' partial
waves, hitting the detector, indeed change the detector's wave
function, but their dynamical relevance to the detector's Bohmian
position is negligible. Thus, according to this picture, it should
be the Bohmian arrival times, which show up in time resolved
detection experiments.

Sections 2 and 3 summarize the basic facts about arrival time
densities according to Kijowski and Bohm. In section 4 we proof
two theorems relating these two notions. The moral of our study is
distilled in section 5. A concise review of Bohmian mechanics can
be found in \cite{Survey}.


\section{Kijowski's arrival time densities}
In 1974 Kijowski \cite{kijowski} introduced a set of conceivable
quantum mechanical arrival time probability densities for a
subspace of right moving wave functions. These densities are
parametrized by quadratic forms of a certain type. We shall
describe them in what follows.

\begin{definition}
\label{quadratic form} Let $\mathcal{D}$ be a complex vector
space. A function $q: \mathcal{D} \rightarrow \mathbb{R}$ is
called a quadratic form, if there exists a hermitian sesquilinear
form $S:\mathcal{D} \times \mathcal{D} \rightarrow \mathbb{C}$
such that $q(\phi) = S(\phi,\phi)$ for all $\phi \in \mathcal{D}$.
\end{definition}

\begin{definition} Let $\mathcal{D}(\mathbb{R}_{+})$ denote the
space of test functions with compact support in
$\mathbb{R}_{+}:=]0, \infty[$ with the usual notion of
convergence. Then
\[
\phi \mapsto \phi_{t} \; \mbox{with} \; \phi_{t}(k)=
\exp\left(-(\rmi \hbar k^2 t /2 m ) \right) \phi(k) \;  \mbox{for}
\; t \in \mathbb{R}
\]
gives the free Schr\"odinger time evolution. Let
$\mathcal{Q}$ denote the set of all continuous quadratic forms $q:
\mathcal{D}(\mathbb{R}_{+}) \rightarrow \mathbb{R}$ such that for
all $\phi \in \mathcal{D}(\mathbb{R}_{+})$ holds

\begin{enumerate}

\item $q(\phi) \geq 0$,

\item $q(\overline{\phi})=q(\phi)$,

\item $\int_{-\infty}^{\infty} q(\phi_{t})\; \rmd t = \| \phi
\|^2$,

\item \label{iv} $\overline{t^2}(q,\phi):=\int_{-\infty}^{\infty}
t^2 q(\phi_{t})\; \rmd t < \infty$.

\end{enumerate}
For any $q \in \mathcal{Q}$ the non-negative function $D_{\phi,
q}: \mathbb{R} \rightarrow \mathbb{R},t \mapsto q(\phi_{t})$
yields a conceivable arrival time density at $x=0$ for the wave
function $\phi \in \mathcal{D}(\mathbb{R}_{+})$ subject to $\|
\phi \|=1$. The case of arbitrary $x \in \mathbb{R}$ is obtained
by replacing $\phi$ in $D_{\phi, q}$ with the function $k \mapsto
e^{-i k x} \phi(k)$.
\end{definition}
According to (\ref{iv}), for all $q \in \mathcal{Q}$ the second
moment of the density $D_{\phi, q}$ is finite. Due to the
continuity of $q$ also the first moment
\[
\overline{t}(q,\phi):=\int_{-\infty}^{\infty}t\, q(\phi_{t})
\;\rmd t
\]
is finite. The variance of arrival times is given by $V(q,\phi):=
\overline{t^2}(q,\phi)-(\overline{t}(q,\phi))^2$. The quadratic
form $q_{0}: \mathcal{D}(\mathbb{R}_{+}) \rightarrow \mathbb{R}$
\[
q_{0}(\phi) := \frac{\hbar}{(2\pi) m} \left|
\int_{-\infty}^{\infty} \sqrt{k} \; \phi(k) \; \rmd k \right|^2,
\]
belongs to $\mathcal{Q}$. The probability density $D_{\phi,
q_{0}}$ is equal to the arrival time density derived from the
Aharonov-Bohm arrival time operator and it is distinguished by the
following theorem.

\begin{theorem}[Uniqueness Theorem]
For all $q \in \mathcal{Q}$ and for all $\phi \in \mathcal{D}
(\mathbb{R}_{+})$ with $\|\phi \|=1$ there holds
\begin{enumerate}
\item \label{equality of t}
$\overline{t}(q_{0},\phi)=\overline{t}(q,\phi)$ and \item
$V(q_{0},\phi)\! \leq \! V(q,\phi)$.
\end{enumerate}
Furthermore $V(q,\phi)=V(q_{0},\phi)$ for all
$\phi\in\mathcal{D(\mathbb{R_{+}})}$ with $\|\phi \|=1$ if and
only if $q=q_{0}$.
\end{theorem}
The proof of this theorem is to be found in \cite{kijowski}.


\section{Bohmian arrival time density}
Let $x \mapsto \Phi_{t}(x):= 1/\sqrt{2\pi}\int_{-\infty}^{\infty}
\exp(\rmi kx) \phi_{t}(k) \, \rmd k$ denote the freely evolving
configuration space wave function at time $t$ associated with the
momentum space wave function $\phi \in
\mathcal{D}(\mathbb{R})\backslash 0$. Let $P_{\phi}(t)$ denote the
probability measure of the set of this wave function's Bohmian
trajectories which cross $x=0$ at some time $s \in ]-\infty, t]$.
Again the arrival at arbitrary $x \in \mathbb{R}$ is obtained by
replacing $\phi$ in $P_{\phi}$ with the function $k \mapsto e^{-i
k x} \phi(k)$. (If we assume an ideal detector to be placed at
$x=0$, then, according to Bohmian mechanics, $P_{\phi}(t)$ is
equal to the detection probability of that wave function at any
time $s \in ]-\infty, t]$.) Let
\[
J_{x}(\phi):=\frac{\hbar}{2m\rmi} (\overline{\Phi(x)}
\frac{\partial} {\partial x} \Phi(x)-\Phi(x)
\frac{\partial}{\partial x} \overline{\Phi(x)} )
\]
denote this wave function's probability current at $t=0$ and at
position $x$. It follows that
\[
\label{J_{0}} J_{0}(\phi)=\frac{\hbar}{2m}\left\{\frac{1}{(2
\pi)}\int_{-\infty}^{\infty}
\int_{-\infty}^{\infty}\!\!(k+l)\,\overline{\phi(l)} \phi(k) \;
\rmd k\:\rmd l\right\} .
\]
Then the following two theorems hold
\cite{kreidl+GG}:
\begin{theorem}
Let $\phi \in \mathcal{D}(\mathbb{R})$ with $\| \phi \|=1$. Then
for all $t \in \mathbb{R}$
\begin{eqnarray}
\label{P_phi} P_{\phi}(t)=\sup\{f_{\phi}(s)|-\infty < s \leq t \}
+\sup\{-f_{\phi}(s)|-\infty< s \leq t\}
\end{eqnarray}
with
\[
f_{\phi} (t):=\int_{-\infty}^{t}J_{0}(\phi_{s})\,\rmd s .
\]

\end{theorem}
From the detection probability one can define in the usual way a
conditional arrival time probability density $B_{\phi}: \mathbb{R}
\rightarrow \mathbb{R}$ by
\[
\frac{P_{\phi}(t)}{\lim_{s \rightarrow \infty}P_{\phi}(s)} =
\int_{-\infty}^{t} B_{\phi}(s) \; \rmd s.
\]
\begin{theorem}
For $\phi \in \mathcal{D}(\mathbb{R})$ and $\|\phi\|=1$ holds
\begin{eqnarray}
\label{allgemeines B} B_{\phi}(t)&=&\left(\lim_{s \to \infty}
P_{\phi}(s)\right)^{-1} \left[J_{0}(\phi_{t})\cdot \chi
\left(f_{\phi}( t)-\sup_{-\infty < s\leq t}\{f_{\phi}(
s)\}\right)\right.  \nonumber \\
&&- \left.J_{0}(\phi_{t})\cdot\, \chi \left(-f_{\phi}(
t)-\sup_{-\infty < s\leq t} \{-f_{\phi}( s)\}\right)\right] \geq
0.
\end{eqnarray}
Here $\chi$ denotes the cutoff function
\begin{eqnarray*}
\chi(s)=\left\{\begin{array}{r@{\qquad for \;\;}l} 0 & s\neq0 \\
1 & s=0.
\end{array}\right.
\end{eqnarray*}
\end{theorem}
The probability current may become negative, even for $\phi \in
\mathcal{D}(\mathbb{R}_{+})$, a fact which is known as the quantum
backflow effect \cite{backflow}. The cutoff function guaranties
the non negativity of the probability density and prevents a
multiple counting of trajectories.

From now on we restrict ourselves to right moving states, i.e.
wave functions $\phi \in \mathcal{D}(\mathbb{R}_{+}) $ with $\|
\phi \|=1$. Due to the half-line localization of $\Phi_{t}$ at
$x<0$ for $t \rightarrow -\infty$ and from probability
conservation we conclude
\begin{eqnarray}
\label{flow} f_{\phi}(t):=\int_{-\infty}^{t} J_{0}(\phi_{s}) \;
\rmd s = \int_{0}^{\infty} | \Phi_{t}(x) |^2 \; \rmd x .
\end{eqnarray}
From this it follows that $0 \leq f_{\phi} \leq 1$. As $\lim_{t
\rightarrow -\infty}f_{\phi}(t)=0$, we have
$\sup\{-f_{\phi}(s)|-\infty< s \leq t\}=0$ for all $t \in
\mathbb{R}$. Thus $P_{\phi}(t)$, according to equation
(\ref{P_phi}), simplifies to
\[
P_{\phi}(t)=\sup\{f_{\phi}(s)|\;-\infty < s \leq t \}.
\]
The half-line localization of $\Phi_{t}$ at $x > 0$ for $t
\rightarrow \infty$ implies $1=\lim_{t \rightarrow
\infty}f_{\phi}(t)=\lim_{t \rightarrow \infty}P_{\phi}(t)$.
Therefore equation (\ref{allgemeines B}) simplifies to
\begin{eqnarray}
\label{spezielles B} B_{\phi}(t)=J_{0}(\phi_{t})\cdot \chi\!
\left(f_{\phi}(t)-\sup_{-\infty < s\leq t}\{f_{\phi}(s)\}\right)
\geq 0.
\end{eqnarray}


\section{Bohm versus Kijowski}
In this section we investigate the question whether the Bohmian
arrival time density $B_{\phi}$ belongs to the class of arrival
time densities considered by Kijowski. This is the case if and
only if there exists a quadratic form $q \in \mathcal{Q}$ such
that
\[
B_{\phi}(t) = q\left( \phi_{t}\right)
\]
for all $\phi \in \mathcal{D}(\mathbb{R}_{+})$ with $\|\phi \|=1$
and for all $t \in \mathbb{R}$. The following theorem demonstrates
that the answer to the above question is no.
\begin{theorem}
\label{disc} There is no quadratic form $q$ on
$\mathcal{D}(\mathbb{R}_{+})$ such that $B_{\phi}(0)=q(\phi)$ for
all $\phi \in \mathcal{D}(\mathbb{R}_{+})$ with $\| \phi \|=1$.
\end{theorem}
\begin{proof}
Let $\varphi \in \mathcal{D}(\mathbb{R}_{+})$ with $\|\varphi\|=1$
such that $B_{\varphi}(0)>0$. A second unit vector $\psi \in
\mathcal{D}(\mathbb{R}_{+})$ is chosen such that $J_{0}(\psi)<0$.
Thus we have
\[
f_{\psi}(0)= \int_{-\infty}^{0}J_{0}(\psi_{t})\;\rmd t <
\sup_{-\infty < s \leq 0 }\{f_{\psi}(s)\}.
\]
From this then follows by means of equation (\ref{spezielles B})
\begin{eqnarray}
\label{Beweis1} B_{\psi}(0)=J_{0}(\psi)\cdot
\chi(f_{\psi}(0)-\sup_{-\infty < s \leq 0 }\{f_{\psi}(s)\})=0.
\end{eqnarray}
Since $\phi \mapsto J_{0}(\phi)$ is a quadratic form its
restriction to a 2D subspace is continuous. Thus the mapping
\[
\xi \mapsto J_{0}(\cos(\xi) \varphi + \sin(\xi) \psi) =: j(\xi)
\]
is continuous on the interval $[0, \pi/ 2]$. Since $j(\pi
/2)=J_{0}(\psi)<0$ there exists a number $\eta \in ]0, \pi/ 2[$
such that $j(\xi)<0$ for all $\xi \in [\eta , \pi / 2]$. In
consequence the mapping
\[
\xi \mapsto B_{\cos(\xi) \varphi+ \sin(\xi) \psi}(0)=: \beta(\xi)
\]
obeys $\beta(0)=B_{\varphi}(0)>0$ and $\beta(\xi)=0$ for all $\xi
\in [\eta, \pi / 2]$.

Assume now that there exists a quadratic form $q \in
\mathcal{D}(\mathbb{R}_{+})$ such that $B_{\phi}(0)=q(\phi)$ for
all $\phi \in \mathcal{D}(\mathbb{R}_{+})$ with $\| \phi \|=1$.
Let $S$ denote the hermitian sesquilinear form associated with
$q$. Then we have
\begin{eqnarray}
\label{q} \beta(\xi)=\cos^2(\xi) q(\varphi) + \sin^2(\xi) q(\psi)
+ \sin(\xi) \cos(\xi) 2 \Re{\left(S(\varphi,\psi)\right)}.
\end{eqnarray}
Since $\beta(0)>0$, equation (\ref{q}) implies
\begin{eqnarray}
\label{assumption} q(\varphi)>0.
\end{eqnarray}
Similarly $\beta(\pi/2)=0$ implies $q(\psi)=0$. Let $\epsilon \in
]\eta, \pi/2[$. Then we have
\[
 0=\frac{\beta(\eta)}{\cos^2(\eta)}=q(\varphi)+2
\Re{\left(S(\varphi,\psi)\right)} \tan(\eta)
\]
and
\[
0=\frac{\beta(\epsilon)}{\cos^2(\epsilon)}=q(\varphi)+2
\Re{\left(S(\varphi,\psi)\right)} \tan(\epsilon).
\]
Since $\epsilon \neq \eta $ and $\tan:[0,\pi/2[ \rightarrow
\mathbb{R}$ is injective we conclude from
\[
2 \Re{\left(S(\varphi,\psi)\right)} \cdot \tan(\epsilon) = 2
\Re{\left(S(\varphi,\psi)\right)} \cdot \tan(\eta)
\]
that $\Re{\left(S(\varphi,\psi)\right)}=0$. Due to
$\beta(\epsilon)=0$ we now have $q(\varphi)=0$ in contradiction to
$q(\varphi)>0$ (see equation (\ref{assumption})).
\end{proof}
Now we compare the first moments of the probability densities
$D_{\phi, q}$ according to Kijowski on the one side, and
$B_{\phi}$ according to Bohmian mechanics on the other side. As
has been shown in \cite{kijowski}, for all $q \in \mathcal{Q}$ and
for all $\phi \in \mathcal{D}(\mathbb{R}_{+})$ with $\| \phi \|=1$
there holds
\begin{eqnarray}
\label{gleichheit} \int_{-\infty}^{\infty} t J_{0}(\phi_{t})\;
\rmd t =\int_{-\infty}^{\infty} t q(\phi_{t}) \; \rmd t=:
\overline{t}(q,\phi).
\end{eqnarray}
In view of the backflow effect this is somewhat surprising. The
following theorem relates the first moments of $D_{\phi, q}$ and
of $B_{\phi}$. The first moment of latter density is denoted as
\[
\overline{t}(B_{\phi}):=\int_{-\infty}^{\infty} t B_{\phi}(t) \;
\rmd t.
\]
\begin{theorem}
Let $\phi$ be in $\mathcal{D}(\mathbb{R}_{+})$ with $\| \phi
\|=1$. Then $\overline{t}(q,\phi) \geq \overline{t}(B_{\phi})$ for
all $q \in \mathcal{Q}$. Equality $\overline{t}(q,\phi) =
\overline{t}(B_{\phi})$ holds if and only if $J_{0}(\phi_{t}) \geq
0$ for all $t \in \mathbb{R}$.
\end{theorem}

\begin{proof}
For $\phi \in \mathcal{D}(\mathbb{R}_{+})$ with $\|\phi \|=1$ and
with $J_{0}(\phi_{t}) \geq 0$ for all $t \in \mathbb{R}$, the
function $f_{\phi}(t)$ is nondecreasing. Therefore, according to
equation (\ref{spezielles B}), there holds
$J_{0}(\phi_{t})=B_{\phi}(t)$ for all $t \in \mathbb{R}$. Thus
from equation (\ref{gleichheit}) we conclude $\overline{t}(q,\phi)
= \overline{t}(B_{\phi})$.

Assume now $\phi \in \mathcal{D}(\mathbb{R}_{+})$ with $\| \phi
\|=1$ and that there exists a $t \in \mathbb{R}$ such that
$J_{0}(\phi_{t})<0$. Then the open set
\[
\Delta_{<}:= \left\{t \in \mathbb{R}\, \Big{|} \,f_{\phi}(t) <
\sup_{-\infty< s \leq t} \{f_{\phi}(s)\} \right\}\subset
\mathbb{R}
\]
is nonempty. Note that for all $t \in \mathbb{R} \backslash
\Delta_{<}$ the equality $f_{\phi}(t)=\sup_{-\infty< s \leq
t}\{f_{\phi}(s)\}$ holds. The set $\Delta_{<}$ is a disjoint union
of open intervals $]a,b[$ such that $f_{\phi}(a)=f_{\phi}(b)$.
Then we have, according to equation (\ref{gleichheit}), that
\begin{eqnarray}
\label{aufteilung_delta} \overline{t}(q,
\phi)=\int_{\mathbb{R}\backslash \Delta_{<}} t J_{0}(\phi_{t})
\rmd t + \int_{\Delta_{<}} t J_{0}(\phi_{t}) \rmd t.
\end{eqnarray}
Equation (\ref{spezielles B}) implies
\begin{eqnarray*}
B_{\phi}(t)=\left\{\begin{array}{r@{\quad}l} 0 & \mbox{for all}
\; t \in \Delta_{<} \\
J_{0}(\phi_{t}) & \mbox{for all}\; t \in \mathbb{R}\backslash
\Delta_{<}
\end{array}\right. .
\end{eqnarray*}
From this and from equation (\ref{aufteilung_delta}) we infer
\[
\overline{t}(q, \phi)=\int_{\mathbb{R}\backslash \Delta_{<}} t
B_{\phi}(t)\rmd t + \int_{\Delta_{<}} t J_{0}(\phi_{t}) \rmd t =
\overline{t}(B_{\phi}) + \int_{\Delta_{<}} t J_{0}(\phi_{t}) \rmd
t.
\]
The latter integral over $\Delta_{<}$ is a sum of integrals over
disjoint intervals $]a,b[$. Denote
\[
F(t)=f_{\phi}(t)-f_{\phi}(a)= \int_{a}^{t}J_{0}(\phi_{s})\rmd s
\]
then holds $F'(t)=J_{0}(\phi_{t})$ and $F(t)<0$ for all $t \in
]a,b[$ and $F(a) = F(b) = 0$. With this each of the integrals can
be estimated by means of a partial integration as follows.
\[
\int_{a}^{b} t J_{0}(\phi_{t}) \rmd t = \int_{a}^{b} t F'(t) \rmd
t = t F(t)|_{a}^{b} - \int_{a}^{b} F(t) \rmd t = -\int_{a}^{b}
F(t) \rmd t
>0.
\]
Thus for wave functions with backflow we have $\overline{t}(q,
\phi) - \overline{t}(B_{\phi})=-\int_{\Delta_{<}} F(t) \rmd t>0$.
\end{proof}
In a completely analogous way one can show for $\phi \in
\mathcal{D}(\mathbb{R}_{-})$ that $\overline{t}(q,\phi) \leq
\overline{t}(B_{\phi})$. See \cite{mine}.


\section{Conclusion}

Now, as the Bohmian arrival time density, associated with a wave
function $\phi \in \mathcal{D}(\mathbb{R}_{+})$, does not belong
to the set of arrival time densities introduced by Kijowski, one
may wonder how deep this discrepancy goes. Indeed, as is obvious
from our proof of Theorem (\ref{disc}), the arrival time density
$B_{\phi}$ violates the quadratic form structure, which is
considered as one of the basic rules of standard quantum
mechanics. To recall this point: We have seen that for some
$\varphi$, $\psi \in \mathcal{D}(\mathbb{R}_{+})$ with $\| \varphi
\|= \| \psi \|=1$ the Bohmian arrival time density obeys
\[
B_{\cos(\xi) \varphi + \sin(\xi) \psi}(0) \neq a+ b \cos(2 \xi) +
c \sin(2 \xi)
\]
for any choice of the constants $a,b,c \in \mathbb{R}$. This fact
also contradicts that version of Bohmian mechanics, which is
empirically equivalent to standard quantum mechanics. The essence
of that version is expressed most succinctly in proposition (2) of
\cite{Survey}. It is crucial for this proposition that the random
variables on the configuration space, which are averaged over with
the position density, are not allowed to depend on the wave
function. The definition of the Bohmian arrival time density
$B_{\phi}$, however, makes use of a random variable, which
parametrically depends on the wave function $\phi$.


\section*{Acknowledgments}

Several ideas for the present work have been received at the
conference ''Quantum theory without Observers II'' in Bielefeld.
The authors thank the organizers for their hospitality, the
stimulating discussion sessions, and financial support. S. Kreidl
in addition thanks for the support by DOC-FFORTE [Doctoral
scholarship program of the Austrian Academy of Sciences].


\end{document}